\documentclass[reprint,amsmath,amssymb,aps,superscriptaddress ]{revtex4-1}

\usepackage{amsmath,amssymb,amsfonts}
\usepackage{algorithmic}
\usepackage{graphicx}
\usepackage{epsf}
\usepackage{epsfig}
\usepackage{textcomp}
\usepackage{xcolor}
\usepackage{float}

\graphicspath{{Fig/}}

\newcommand{\blue}{\color{black}}    % change blue text to black

\begin{document}

\title{Effect of Impurities on Charge and Heat Transport in Tubular Nanowires}

\author{Hadi Rezaie Heris}
\affiliation{Department of Engineering, Reykjavik University, Menntavegur 1, IS-102 Reykjavik, Iceland}

\author{Kristjan Ottar Klausen}
\affiliation{Department of Engineering, Reykjavik University, Menntavegur 1, IS-102 Reykjavik, Iceland}

\author{Anna Sitek}
\affiliation{Department of Theoretical Physics, Wroclaw University of Science and Technology, Wybrze\.{z}e  Wyspia\'{n}skiego  27, 50-370 Wroclaw, Poland}

\author{Sigurdur Ingi Erlingsson}
\affiliation{Department of Engineering, Reykjavik University, Menntavegur 1, IS-102 Reykjavik, Iceland}

\author{Andrei Manolescu}
\affiliation{Department of Engineering, Reykjavik University, Menntavegur 1, IS-102 Reykjavik, Iceland}

\begin{abstract}

{\blue We calculate the charge and heat currents carried by electrons, originating from a temperature gradient and a chemical potential difference between the two ends of tubular nanowires with different geometries of the cross-sectional areas: circular, square, triangular, and hexagonal. We consider nanowires based on InAs semiconductor material, and use the Landauer-B\"{u}ttiker approach to calculate the transport quantities. We include impurities in the form of delta scatterers and compare their effect for different geometries.  The results depend on the quantum localization of the electrons along the edges of the tubular prismatic shell. For example, the effect of impurities on the charge and heat transport is weaker in the triangular shell than in the hexagonal shell, and the thermoelectric current in the triangular case is several times larger than in the hexagonal case, for the same temperature gradient. }

\end{abstract}

\maketitle

\section{Introduction}
{\blue Down to the nanoscale, the closeness of constituting atoms to material boundaries and
high surface area to volume ratio provides interesting quantum physical phenomena. Nanostructures exhibit unique properties that are different from those of their bulk counterparts. Among these nanostructures, nanowires have
attracted considerable attention due to their potential applications in modern technologies. Nanowires of small size are ideal for miniaturized devices such as transistors \cite{duan2003high}, batteries \cite{park2007preparation}, and sensors \cite{fakhri2022germanium,fakhri2022piezoresistance}. Also, their high surface area to volume ratio increases the efficiency of energy storage, and provides unique physical and chemical properties such as improved electrical conductivity, increased strength, and enhanced catalytic activity \cite{zhou2019nanowires,yu2018nanowires,wang2013silicon}. The properties of nanowires depend on their material, as they can be made from metals, semiconductors, and insulators, but also on their internal geometry. } 

Semiconductor nanowires are potential candidates for several fields of technology such as nanoelectronics \cite{xu2010self,lu2007nanoelectronics,bansal2020highly,alexsandro2021low,yuan2021reducing}, quantum information processing \cite{hu2007ge}, solar cells \cite{kempa2008single,gao2022highly}, biological sensors \cite{cui2001nanowire},  thermoelectrics and energy conversion devices \cite{boukai2008silicon,hochbaum2008enhanced,li2021si}, and integrated circuits \cite{nam2009vertically}. In many of these applications understanding the charge and heat current is crucial. Thermoelectric devices demand a high charge current associated with low heat transport to reach high efficiency \cite{snyder2008complex,peri2022giant, heris2020thermoelectric}, while nanoelectronic devices necessitate high heat transport and sinks to take heat out from nano chips \cite{schelling2005managing}. In particular, core-shell nanowires, which are radial heterojunctions of two or more different semiconductor materials, enable the control of charge and heat transfer through specific geometry. In such structures, electronic properties can be determined by the band alignment between materials, core size, and shell thickness. Similarly, heat transport can be guided or trapped through core-shell nanowires \cite{heris2022effects}. 
 
 Core-shell nanowires with a doped shell and an undoped core are tubular conductors, such that conduction takes place only in the shell \cite{gul2014flux}. It is also possible to achieve tubular nanowires by etching the core part \cite{haas2013nanoimprint}. {\blue Tubular nanowires, with length much greater than their diameter, have special mechanical, electrical, and optical properties, than ordinary full body nanowires. Tubular nanowires can maintain their shape and structure even under extreme conditions, such as high temperatures or pressures \cite{noel2014physical,li2022structural}. Also, high electrical conductivity due to the elimination of phonon scattering, and high chemical stability are other advantages of tubular nanowires \cite{liu2010metal,skakalova2006electronic,adler2006thermal,chunfeng2007chemical}. }

 Semiconductor core-shell nanowires based on III-V materials are most commonly prismatic. The typical shape of the cross-section is hexagonal \cite{blomers2012realization,petronijevic2017chiral,chen2002nanowires}, but other prismatic shapes can also be fabricated, like square \cite{fan2006single} or triangular \cite{heurlin2015structural,khosravi2020torsional}. The prismatic geometry of core-shell nanowires, especially tubular nanowires, presents a unique window to important features. These include conductance  and electron localization \cite{torres2018conductance,sitek2015electron},  %can be determined by internal geometry
quantum confinement effect over charge carriers \cite{tian2012one}, interacting several Majorana states with each other \cite{manolescu2017majorana,klausen2020majorana} and inducing the sign reversal of the electric current generated by the temperature gradient in presence of a transversal magnetic field \cite{erlingsson2017reversal}. {\blue Amongst different materials for prismatic nanowires, InAs is one the most studied.  Different approaches for controlling grow \cite{mandl2006free,tomioka2008control} and lattice \cite{dick2010crystal} features,  ultrahigh electron mobility \cite{ford2009diameter}, and direct-narrow band gap \cite{liu2013high}, are some of the advantages of this material. }

In tubular nanowires apart of the geometry mainly three factors lead to changes in charge and heat current: surface roughness, impurities and phonons. Surface roughness leads to a reduction in charge current but practically it is easy to estimate the range of this deviation for different materials \cite{ziman2001electrons}. {\blue It had been shown that electron-phonon interaction can be neglected in thin shells \cite{choi2003nonlinear}. Also, experimentally it is possible to heat up just electrons, while the phonons are kept frozen \cite{van1992thermo, molenkamp1990quantum}. Consequently, the electron-phonon coupling can be assumed to be negligible.} Impurities play a significant role in the conduction feature of tubular nanowires, and it is different for transported charge and heat current for each cross-section with respect to chemical potential bias or temperature gradient. Different numbers of impurities or varying intensities show distinct behavior in each case of cross-section geometry. Understanding the combination of heat and charge current is essential in many cases of nanoscale studies, and the purpose of this work is to provide a comprehensive view of impurities' effect on electronic conduction of tubular nanowires.  We study the effect of the density of the impurities, and of the strength of the associated potential, on the charge and heat currents, in prismatic shells. 
%Then we will choose two of the promising cross-sections to study more about impurities effects. 
Due to the fact that nowadays it is feasible to fabricate tubular nanowires with the desired cross-section and different doping variations, our study can be useful to establish a direct relation between different cross-sections and transport properties of core-shell  nanowires.
 
The paper is structured in the following sections: Section 2 contains our model and methodology. In Section 3 we present the results, in particular we show how the presence of impurities affects conductance through wires of different cross sections. Further, we focus on wires having hexagonal and triangular cross sections and we study the impact of the number and strength of impurities on the conductance. Finally, Section 4 contains the conclusions.      
 
 %{\sout{Section 2 details our model and methodology. In section 3, results are presented and discussed for the effect of impurities for different cross-sections, the number of impurities effect in a triangular shell, and the intensity of the impurities effect in a triangular shell. Finally, we summarize the conclusions in section 4.}}

\section{Model and methods}

\begin{figure}
    \centering
    \includegraphics[width=0.99\linewidth]{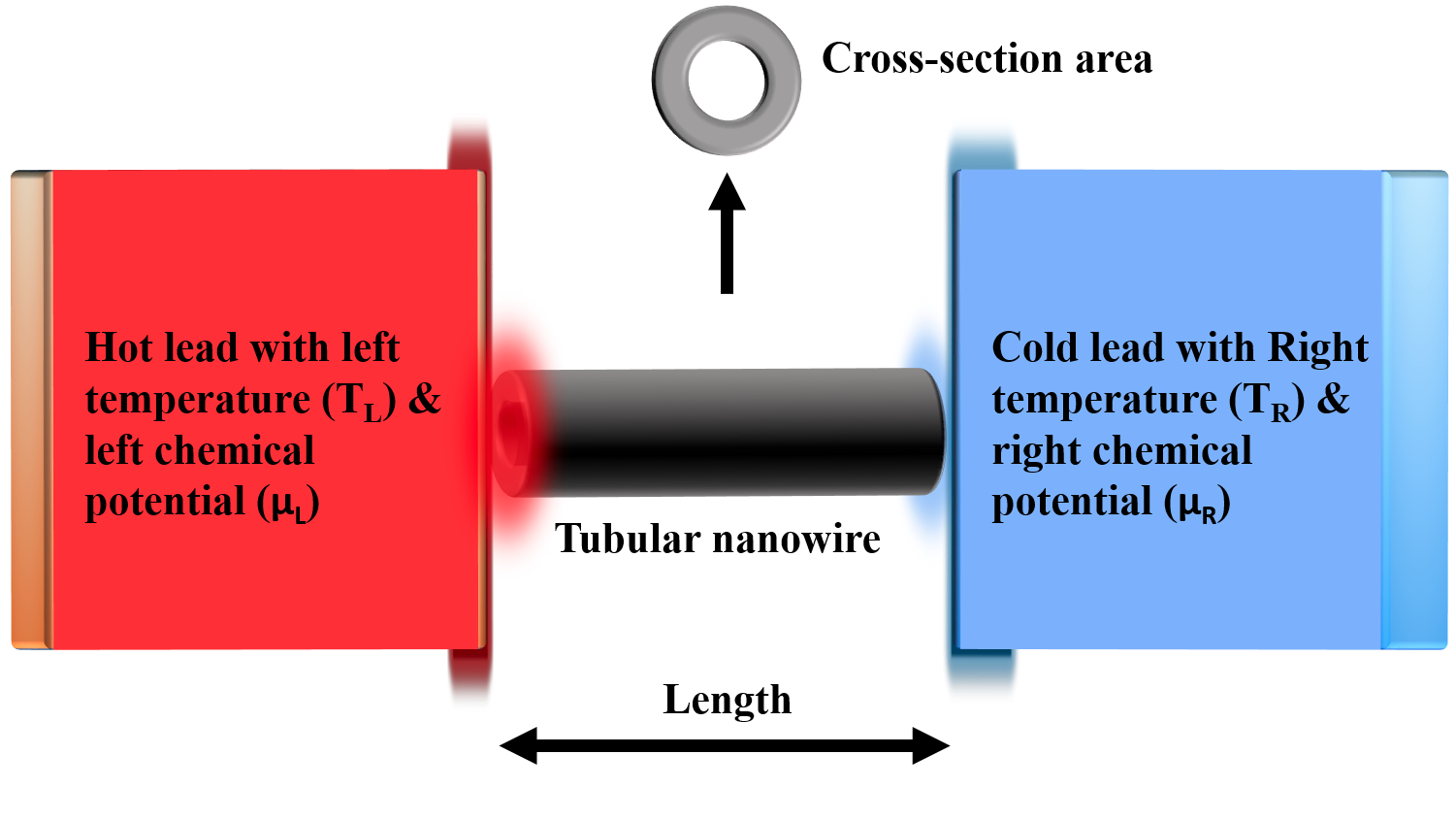}
    \caption{Schematic structure of a tubular nanowire with a circlular cross-section in the presence of a temperature gradient and a chemical potential bias between hot and cold leads.}
    \label{schematic}
\end{figure}

Our model is {\blue a core-shell nanowire} with a finite length and with different cross-sectional geometries. 
{\blue The different shapes of the interface between the core and the shell lead to different energies for each electronic state. Consequently, the physical properties of the nanowire will depend on the geometry of its cross section. In the present work we consider the core material not doped, i.\ e. it behaves like an insulator.  The only role of the core is to define the geometry of the shell.  We assume the conduction of electrons occurring  solely through the shell, which is assumed n-doped.} 
We consider a temperature gradient and different chemical potentials between the two ends or the nanowire, {\blue by assuming it placed between two macroscopic leads, as illustrated in Figure \ref{schematic}.}

{\blue The disorder is represented by localized repulsive potential barriers, mimicking the effect of impurities present in the nanowire. All impurities are assumed of identical strength, and also fixed and rigid. They are randomly distributed within the volume of the shell, using a random number generator. The idea behind this kind of impurity model is that all physical effects on the electronic conduction occur due to the sequential scattering of an electron on individual impurities, when the distance between the electron and impurity is very short, and then the electron propagates freely until eventually meeting the next impurity, where another scattering event occurs.}
Then for each example, we consider a particular number of impurities whose strength varies within a certain range. 

\begin{figure}%[H]
  \centering
  \includegraphics[trim = 0 16 215 45, clip, width=0.48\linewidth]{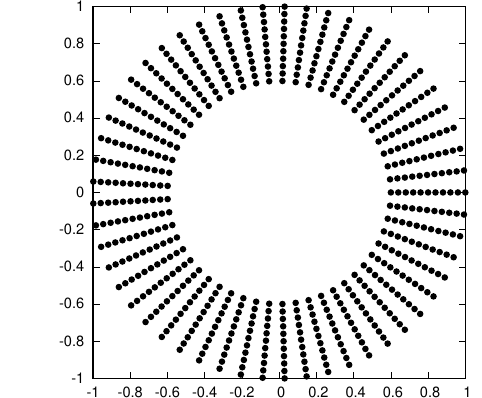} 
  \includegraphics[trim = 0 16 215 45, clip, width=0.48\linewidth]{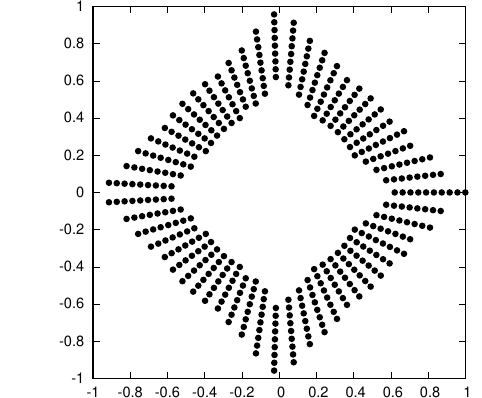}
  \includegraphics[trim = 0 16 215 30, clip, width=0.48\linewidth]{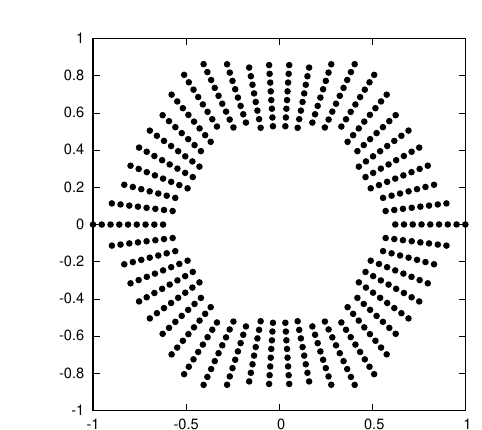}
   \includegraphics [trim = 23 13 53 0, clip, width=0.48\linewidth]{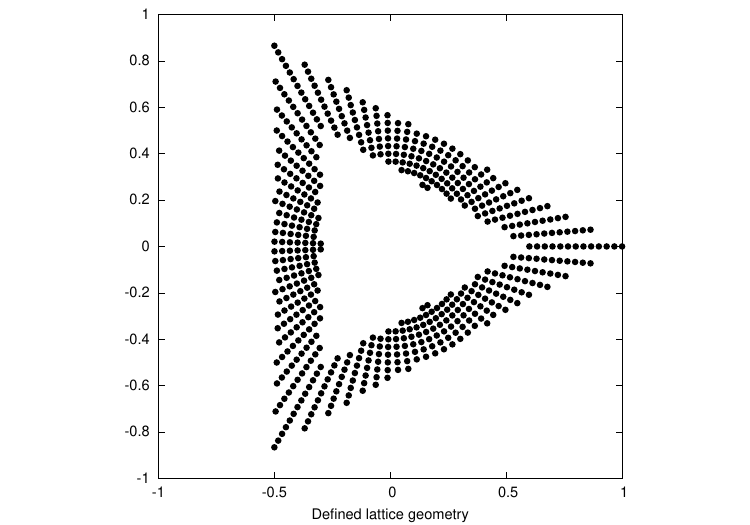}
  \caption{The discretized polygonal with external radius of 50\,nm and thickness of 20\,nm. {\blue The axes correspond to the $x$ and $y$ coordinates divided by the radius.} The cross-section of the prismatic shells are defined by applying boundaries on a circular ring discretized in polar coordinates-points indicate the shell thickness. Note that for the numerical calculations all shells consist of a higher number of radial and angular sites then what is shown in the figures, where, for illustration purpose, we used a smaller number of points.  
  }
\label{fig7}
\end{figure}

We calculate the electronic charge and heat currents for four cross section shapes: cylindrical, hexagonal, square and triangular, as shown in Figure \ref{fig7}. In all cases the external radius of the nanowire is 50 nm (i.e. the radius of the circle encompassing the entire cross section) and 
the shell thickness is $20$\,nm, so that conduction takes place in the narrow shell cross-sectional area. All nanowires have a length of 500\,nm. {\blue The number of occupied quantum states in the nanowire, which is also the number of electrons participating to the transport, is up to about 10000.  In terms of carrier density, it corresponds to a doping level of the order of $10^{15}\  {\rm cm}^{-3}$, but in experimental nanowires the carrier concentration is also controlled with electrostatic gates.}

The Hamiltonian of the nanowire consists of a longitudinal and a transverse term. The transverse Hamiltonian is discretized in polar coordinates \cite{daday2011electronic,sitek2016multi}. A specific cross-section of the polygonal shell, defined on a set of lattice of points, is obtained from a circular disk, after excluding every point from the lattice which stands outside the shell boundaries \cite{torres2018conductance}. 
 The electron effective mass is $m_{\rm eff}=0.023\, m_{\rm e}$, as for bulk InAs.
We calculate heat current $I_{q}$ and electrical current $I_{c}$ driven by the temperature gradient and chemical potential bias applied between two ends of nanowire using the Landauer-Buttiker approach:

\begin{equation}
    I_{c}=\frac{e}{h} \int {\cal T}(E)[f_{R}(E)-f_{L}(E)]dE \ ,
    \label{eq2}
\end{equation}
\begin{equation}
    I_{q}=\frac{1}{h} \int {\cal T}(E)[E-\mu][f_{R}(E)-f_{L}(E)]dE \ ,
    \label{eq3}
\end{equation}
where ${\cal T}$ is the transmission function, and
\begin{equation*}
    f_{L,R}(E)=\frac{1}{1-e^{(E-\mu_{L,R})/kT_{L,R}}}
\end{equation*}
is the Fermi function for the left (L) or right (R) reservoir with
chemical potential $\mu_{L,R}$ and temperature $T_{L,R}$ (and $k$ denoting Boltzmann's constant). 

%For a better understanding of occurring phenomena and to make our result clear we study the 
For clarity reasons we study the 
conduction under the temperature gradient case and chemical potential bias case separately. So in each case one type of bias is constant and the other one is variable. 
The chemical potential bias $\Delta \mu=\mu_{L}-\mu_{R}$ is always present %at $3$\,meV 
at the two ends of the nanowire, and in each step of the calculation we change the left and the right reservoir values simultaneously, but keeping $\Delta \mu = 3$\,meV fixed. In all cases, we 
consider the first 10 transverse modes in our numerical calculation regardless of geometries or impurities variation. 

Different cross-sectional areas lead to different energy spectra for each shape and consequently different chemical potential windows appear. Due to this reason, first we need to find the maximum and minimum possible values of the chemical potentials, corresponding to each specific cross-section. 
Then, the entire energy interval between the lowest and highest energy states (corresponding to the first 10 transverse modes) is explored step-wise.
During this series of calculations, i.\ e. in the presence of a chemical potential bias, we consider no temperature gradient, and $T_{R}=T_{L}=200$\,K. To compute the currents, the energy limits are considered between the ground state and the highest available state for each cross-section. For instance, this range for circular cross section is 35-85\,meV, and for triangle cross section is 120-220\,meV.

In the case of a temperature gradient, we fix $\Delta T=T_{L}-T_{R}=35$\,K. In the first step, we set $T_{L}=36$\,K and $T_{R}=1$\,K and then we increase the temperature of both sides simultaneously. This process  is repeated several times, while the left temperature varies between  36 - 420\,K. The temperature gradient effect on the conductance is calculated using the same chemical potentials at the left and right end of the nanowire. For each geometry these values are calculated as $\mu_{L}=\mu_{R}=(\mu_{\rm max}+\mu_{\rm min})/2$,  where the miniminum and maximum values are equal to the minimum and maximum energy in the computed electronic spectra, respectively. For example, in the case of the circular cross section $\mu_{\rm max}=85$\,meV and  $\mu_{\rm min}=35$\,meV, and we set both chemical potentials at $60$\,meV. By applying the same rule to the other geometries, chemical potential values are $75$\,meV, $110$\,meV, and $170$\,meV, for the hexagon, square and triangle, respectively.

%\vspace{2cm}

\section{Results and discussion}

\subsection{Impurities in different cross section geometries }

Tubular nanowires increase thermoelectric current due to the lateral electron confinement \cite{balaghi2021high,ramayya2007electron} and decrease heat transport due to the strong suppression of phonons with a diameter below the phonon mean free path \cite{heris2022effects,juntunen2019thermal}. Core-shell nanowires with polygonal cross-sections show  a stronger electron localization at the corners than on the sides of the polygon.
The energy  structure of shells with polygonal cross-sections has a strong dependence on the number of corners \cite{estarellas2015scattering,sitek2015electron}. There is a remarkable energy gap between the states localized at corners and the next states which stand on  polygon's sides. This gap increases with decreasing the shell thickness or the number of corners \cite{heris2022charge}.
The lowest energy states are always localized in the corners. The polygonal shell properties differ considerably from each other because of the complexity of the localization of low energy states. 

\begin{figure}%[H] 
\centering
\includegraphics[width=1\linewidth]{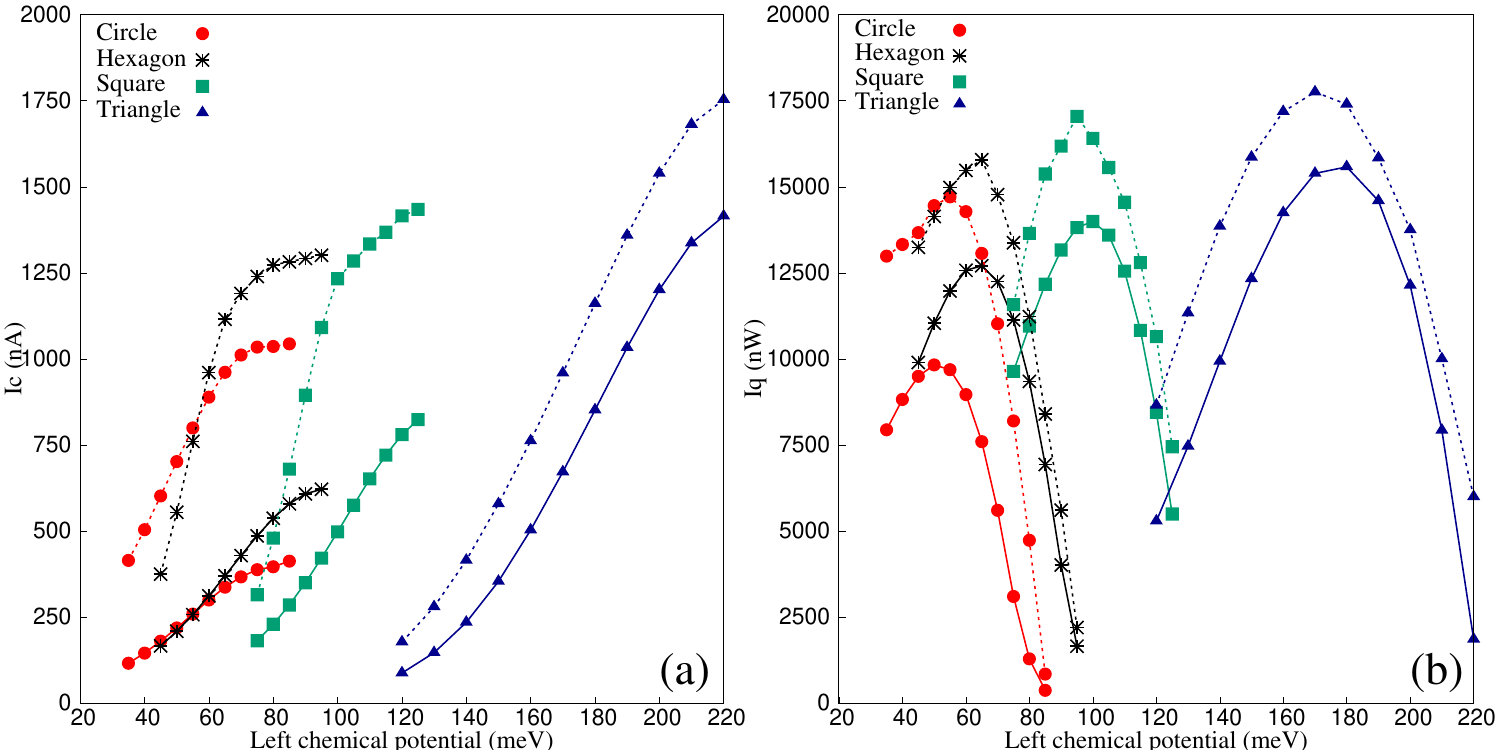}
\includegraphics[width=1\linewidth]{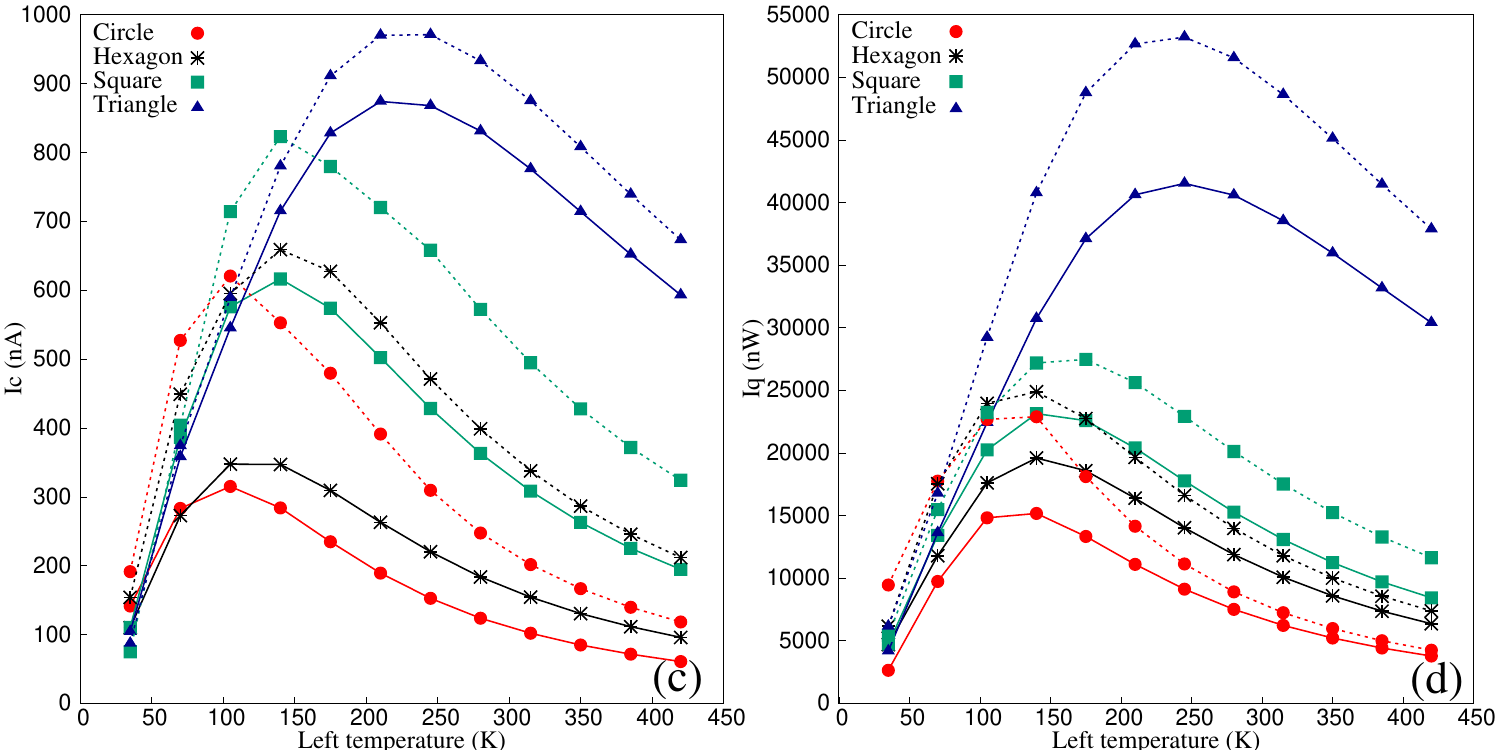}
 \caption{The effect of a high number of impurities, for different geometries of shell, on the charge and heat currents. (a) and (c) figures show the electrical current, and (b) and (d) show the heat current. Conduction features are shown as a function of left chemical potential in (a) and (b), and as a function of left temperature in (c) and (d). The dashed lines represent the tubular nanowires with no impurity and the solid lines represent the nanowires with 10000 impurities each one of 10\,meV. }
\label{shape-number} 
\end{figure}

Implementing disorder in the shell such as impurities  makes these properties even more complex, but at the same time with some promising results. Figure~\ref{shape-number} presents charge and heat current   versus the left chemical potential and temperature for tubular nanowires with different cross-sections in the presence of $10^4$  impurities. All implemented impurities are repulsive with the strength of 10\,meV.
Despite different energy states for each cross-section, the charge current for all geometries follows the transmission function and the energy window associated with the chemical potential. So increasing  the chemical potential allows a larger number of states to participate in the transmission, which leads to a higher charge current.  This occurs for pure wires, and for those with impurities as well.

\begin{figure}%[H] 
\centering
\includegraphics[width=1\linewidth]{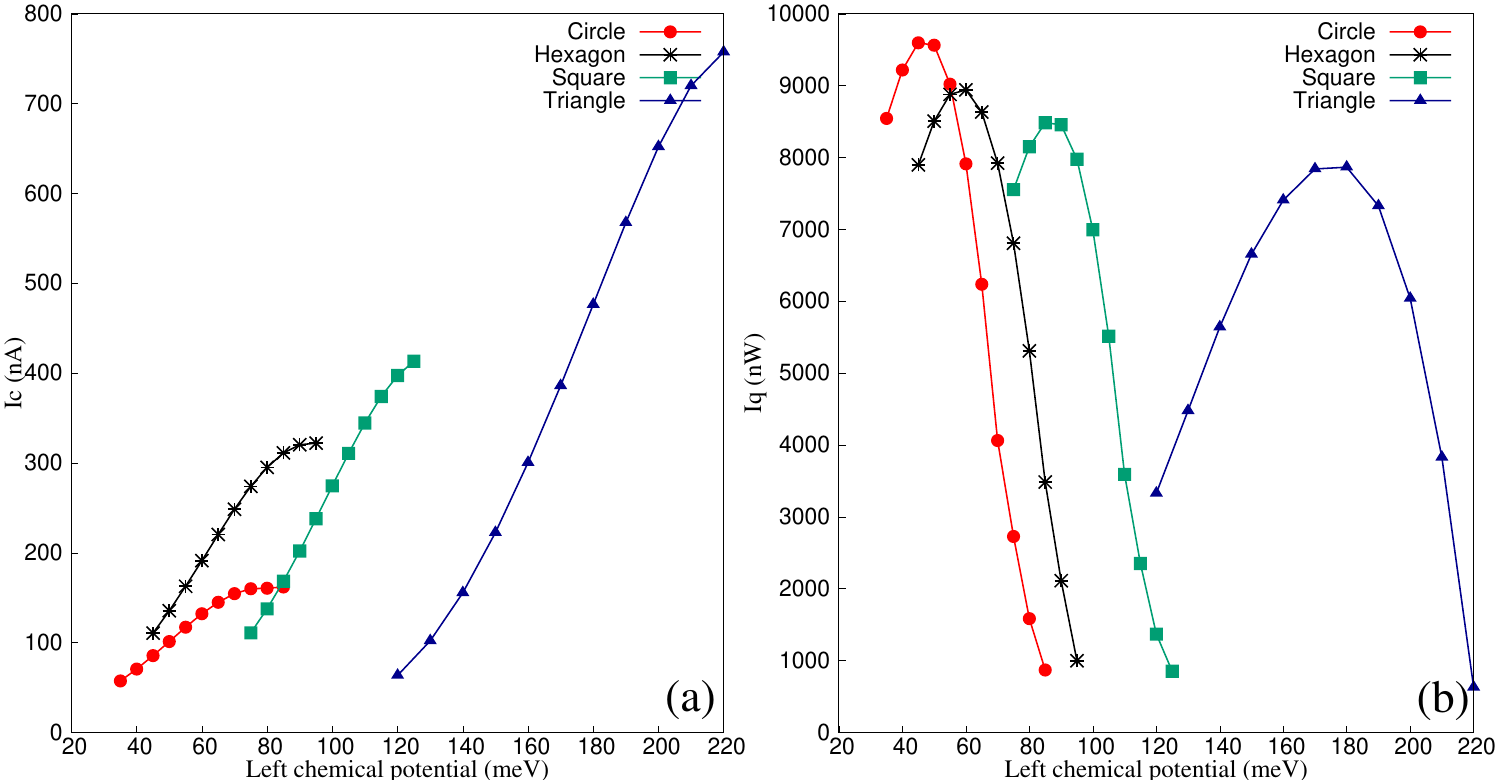}
\includegraphics[width=1\linewidth]{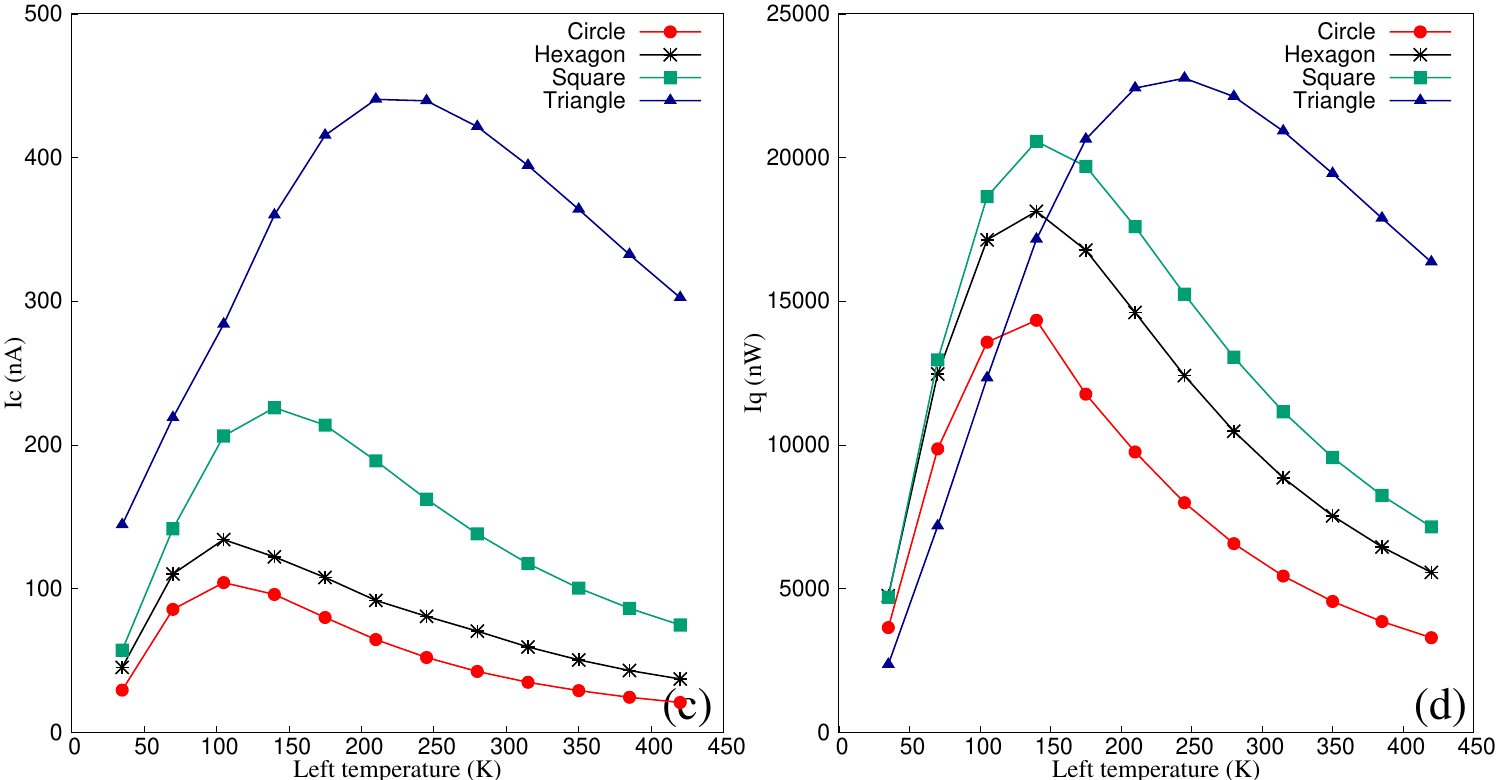}
 \caption{Effect of impurities with strong associated potential on the electric and heat currents for different cross sections of the shell.  Conduction features are studied as a function of left chemical potential in (a) and (b), and as a function of left temperature in (c) and (d). (a) and (c) figures show the electrical current, and (b) and (d) show the heat current.}
\label{shape-strength} 
\end{figure}

It is clear from all curves in Figure~\ref{shape-number} that the presence of impurities leads to a reduction in charge and heat current values with respect to temperature or chemical potential. These reductions can be seen easily as difference between the magnitude of currents with no impurities (dashed lines curves) and the magnitude with impurities (solid lines curves) with the same color.
In Figure~\ref{shape-number}(a) we can see that despite having the same area and number of impurities (10000 impurities with a magnitude of 10\,meV) for all polygonal shapes, the triangular shell can carry charge way better than others. The cross-sections with fewer corners have higher values of thermoelectric current. This can be explained by the fewer transverse modes, but with steeper energy dispersion at the energy reached by higher values of of the variable chemical potential. While in the case of heat current, the variation with the nanowire shape is smaller.

{\blue The maximum heat current occurs in the absence of impurities and above 100\,K.  The location of this maximum varies with respect to the geometry of the nanowire, and it decreases by increasing the number of impurities. Using the cross sectional area $A$ of the nanowire it is possible to calculate the thermal conductivity $\kappa$. For instance, by considering the heat current $52~\mu$W at 250\,K and $\kappa= \frac{Q L}{A \Delta T}$, our thermal conductivity is approximately 26 W/mK. Some experimental studies present comparable results for the thermal conductivity of InAs nanowires \cite{zhou2011thermal,ren2014synthesis}.}

We can see an order between different cross sections in charge and heat transported by electrons as a function of both left chemical potential and left temperature.  Also, the reduction of currents due to impurities is different for each geometry. By using the same number of impurities in different cross-sections, we can see that in the case of the circular cross-section the charge and heat transport decrease dramatically in comparison to other geometries. On the average the circular cross-section shows 65\% reduction of the charge current values and 30\% reduction of the heat current values in the presence of impurities. The corresponding values for the triangular cross-section are 10\% and 15\% , respectively (see Figure~\ref{shape-number}(a) and (b)). Figure~\ref{shape-number}(c) and (d) also present the values of charge and heat current for pure shells and their reduction in the presence of impurities with respect to the left side temperature. These reductions are smaller than in the case of a chemical potential bias. Despite the presence of the $10^4$ impurities, with the magnitude of 10\,meV, the triangular shell still shows the highest  values of both electrical and thermal currents. 

Considering impurities in a system as a disorder, one should take into account both the number and strength of implemented impurities. Therefore, in the next series of calculations instead of the high number of impurities we consider a small amount of them, but with stronger potentials, to see the variation of thermoelectric properties again as a function of chemical potential and temperature for different shell cross sections. In all cases, we implement 300 impurities with a strength of 200\,meV.
According to Figure~\ref{shape-strength}(a) we can see a again significant reduction in the magnitude of both electrical and heat current for all geometries. But interestingly, implementing impurities with strong potentials leads to a reduction of near 70\% for the circular cross-section and 48\% in the case of the triangular cross-section. The reduction of the heat transport for the circular cross section is much weaker than for triangular one, which results in the reversed order of the curves. Compare Figure~\ref{shape-number}(a) and Figure~\ref{shape-strength}(a) for $I_{c}$ values, and compare Figure~\ref{shape-number}(b) and Figure~\ref{shape-strength}(b) for $I_{q}$ values. This noticeable shift in $I_{c}$ and $I_{q}$ magnitudes and sequence order is not limited to chemical potential bias and we can see the same behavior also in the case of the temperature gradient.

\subsection{Effect of the number of impurities}

\begin{figure}%[H] 
\centering
\includegraphics[width=1\linewidth]{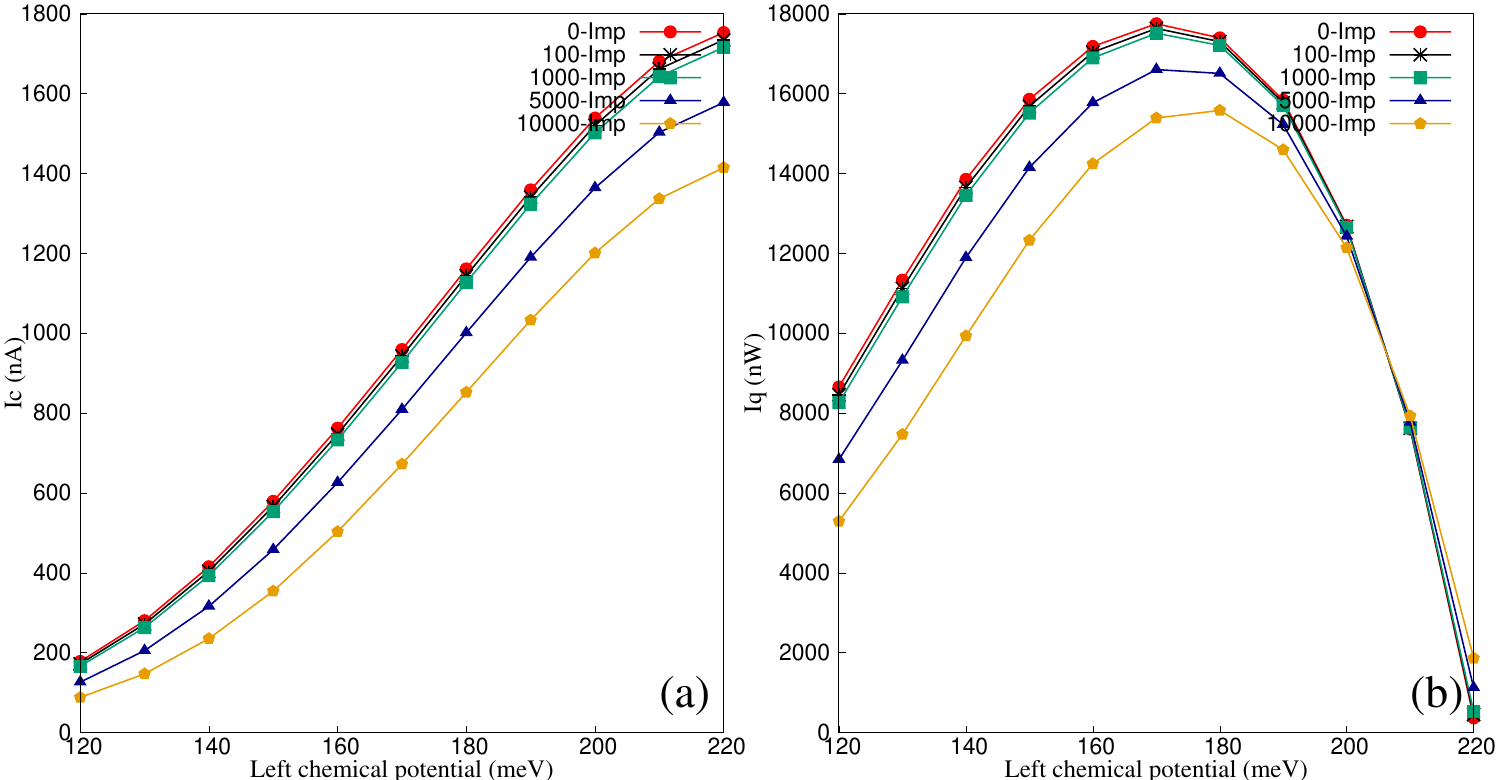}
\includegraphics[width=1\linewidth]{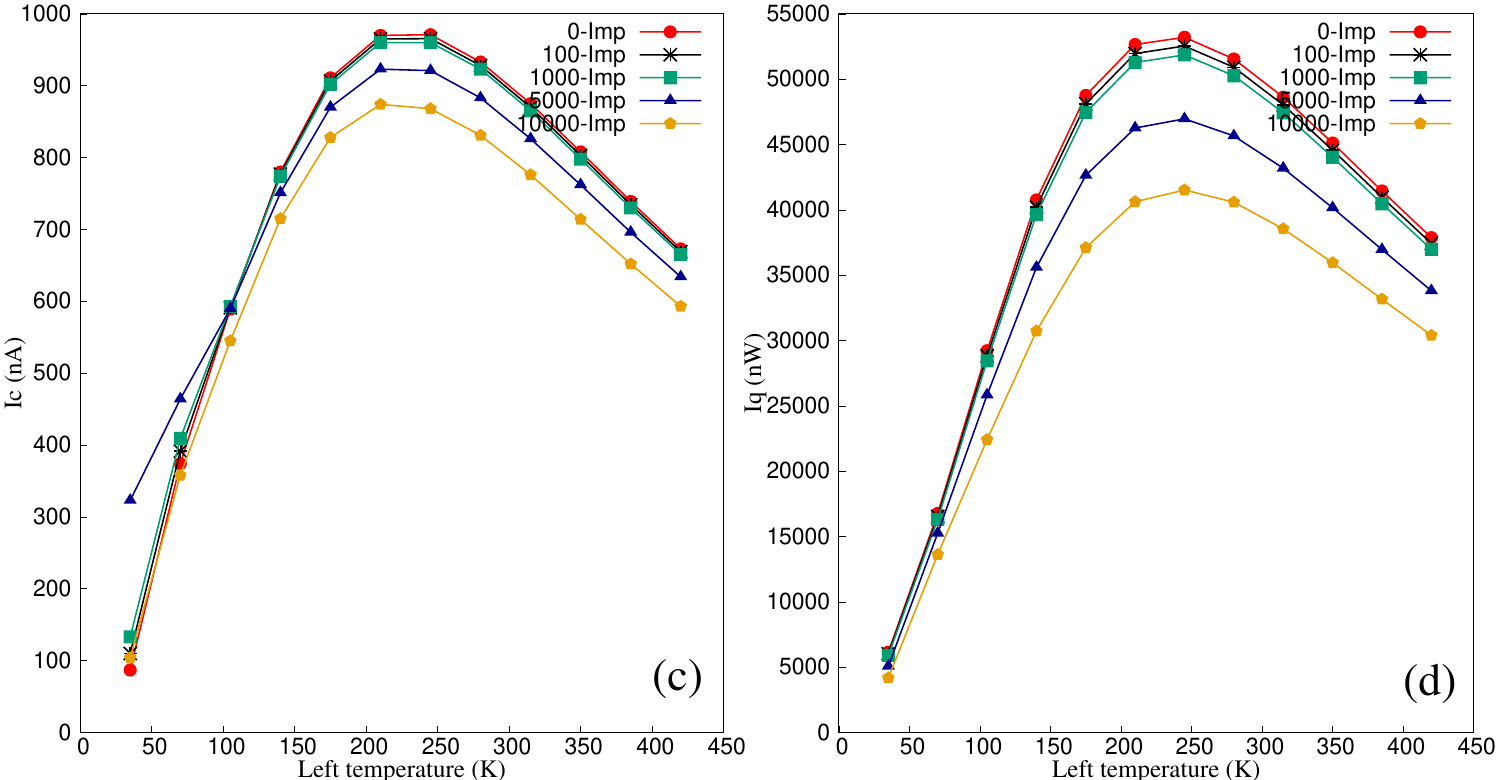}
 \caption{Electronic conduction variation in the triangular shell in the presence of different numbers of impurities. (a) Electrical current as a function of the left chemical potential. (b) Heat current as a function of left chemical potential. (c) Electrical current vs. left temperature. (d) Heat current vs. left temperature.}
\label{number} 
\end{figure}

{\blue Figures \ref{shape-number} and \ref{shape-strength} present the effect of a high number of impurities and of impurities with strong associated potentials
on the electronic properties of nanowires with different cross sections. Results point out
that nanowires with triangular cross-sections show the highest electrical current with respect to
both temperature gradient and chemical potential bias. Also, the heat current has the largest variation for triangular nanowires, compared to the other geometries, in presence of impurities.} 
Due to these reasons, and based on the fact that the typical shape of the cross section for III-V materials is hexagonal, we choose the triangular and hexagonal shells to explore further the effects of impurities in these cases.   
We consider in our systems impurities with different numbers for each case (0, 100, 1000, 5000, 10000) with a fixed magnitude of 10\,meV. Increasing the number of impurities always leads to a reduction in values of both electrical current and heat current regardless of chemical potential or temperature gradient variation. 

\begin{figure}%[H] 
\centering
\includegraphics[width=1\linewidth]{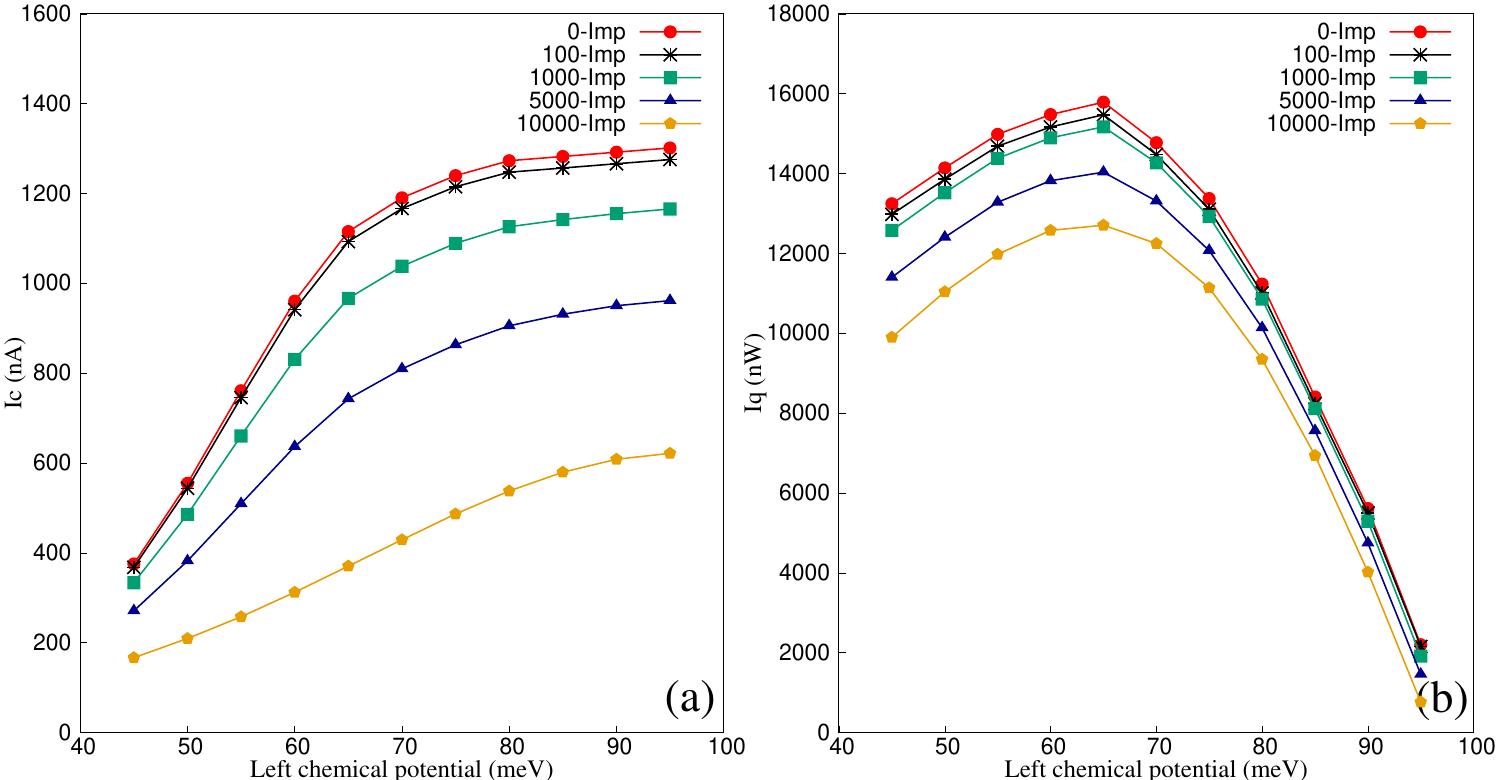}
\includegraphics[width=1\linewidth]{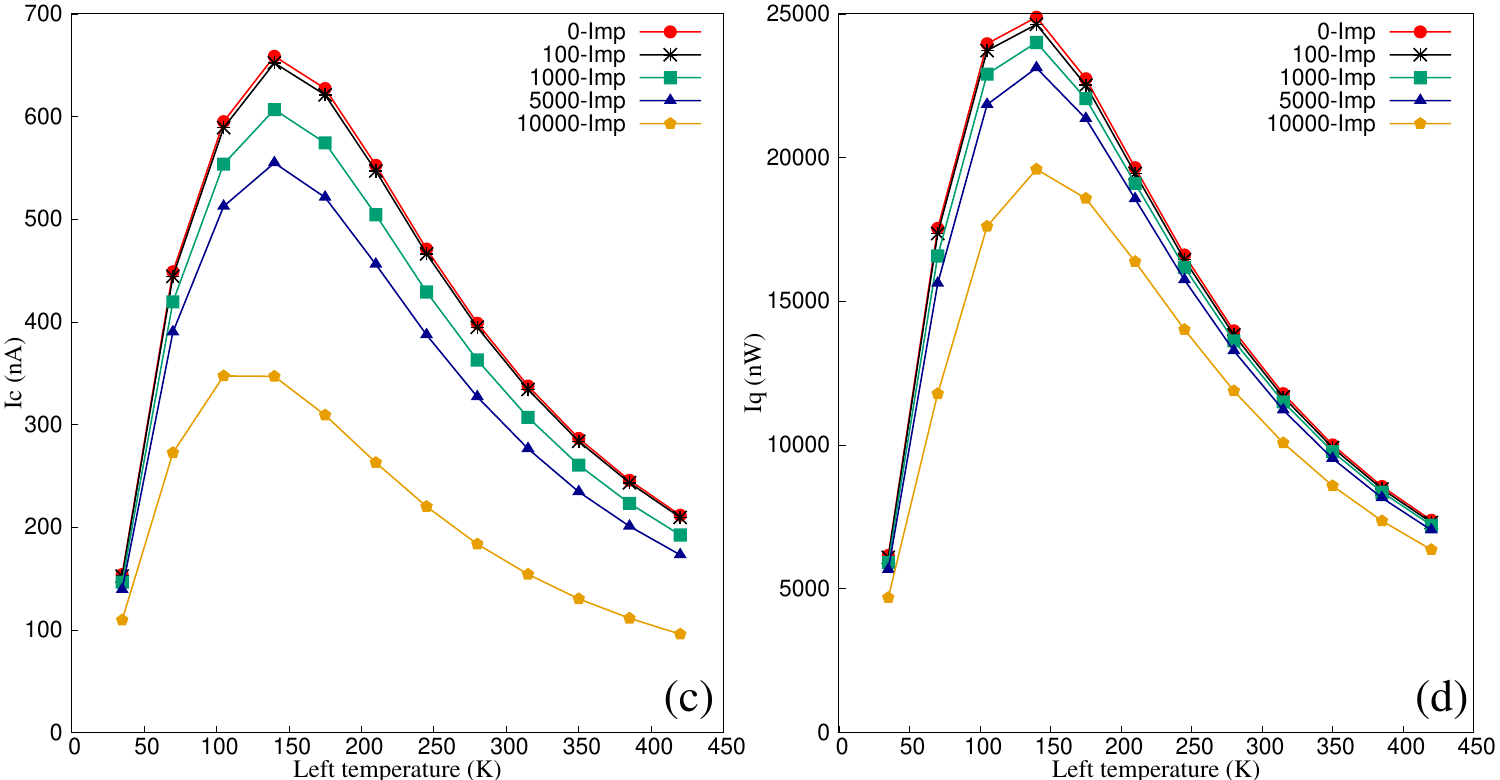}
 \caption{The effect of the number of impurities in the hexagonal shell on the charge and heat
current. The number of impurities in the nanowire is shown in the legend. (a) and (c) figures show the electrical current, and (b) and (d) show the heat current. Conduction features are studied as a function of the left chemical potential in (a) and (b), and as a function of the left temperature in (c) and (d).
}
\label{number-hex} 
\end{figure}

A small number of impurities will not play a 
noticeable role in the reduction of conduction. As can be seen in Figure~\ref{number} (a) and (c) the presence of up to 1000 impurities does not affect much the charge current, but further increase of the number of impurities results in the reduction of 15-25\%  in the current for the triangular shell. The current reduction due to impurities is much larger for hexagonal shells, where the percentage drop may be twice larger than in the case of triangular wire, Figure~\ref{number-hex}. The process that causes the current reduction here is scattering, each impurity acts as a scattering center, and thus the effect increases with the number of impurities in the wire. However, the stronger localization of electrons at corners in the triangular geometry, where the corners are sharper than in the hexagonal case, leads to a more robust electronic states in the triangular case, and thus to currents less sensitive to impurities.

%\newpage

\subsection{Effect of the strength of impurities} 

\begin{figure}%[H] 
\centering
\includegraphics[width=1\linewidth]{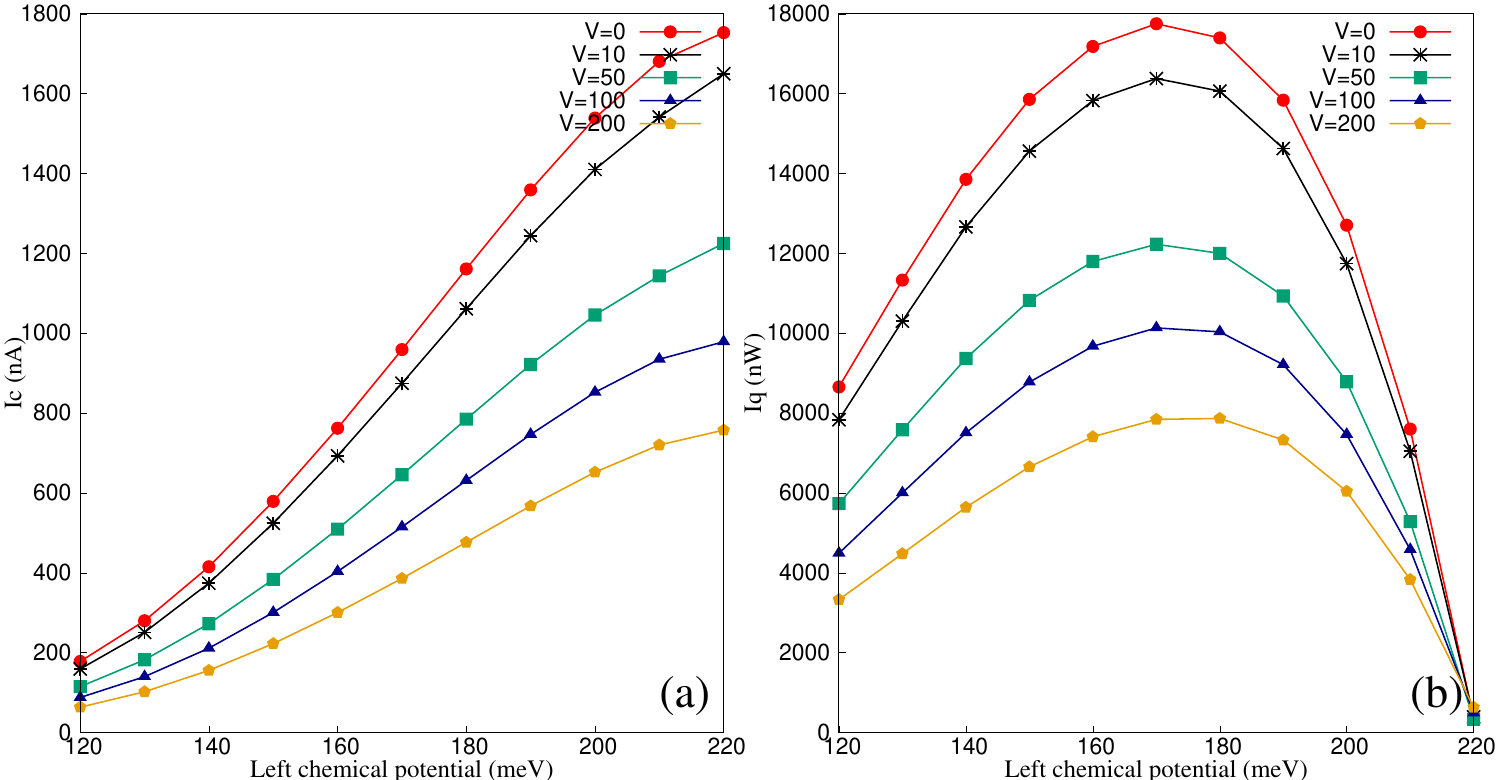}
\includegraphics[width=1\linewidth]{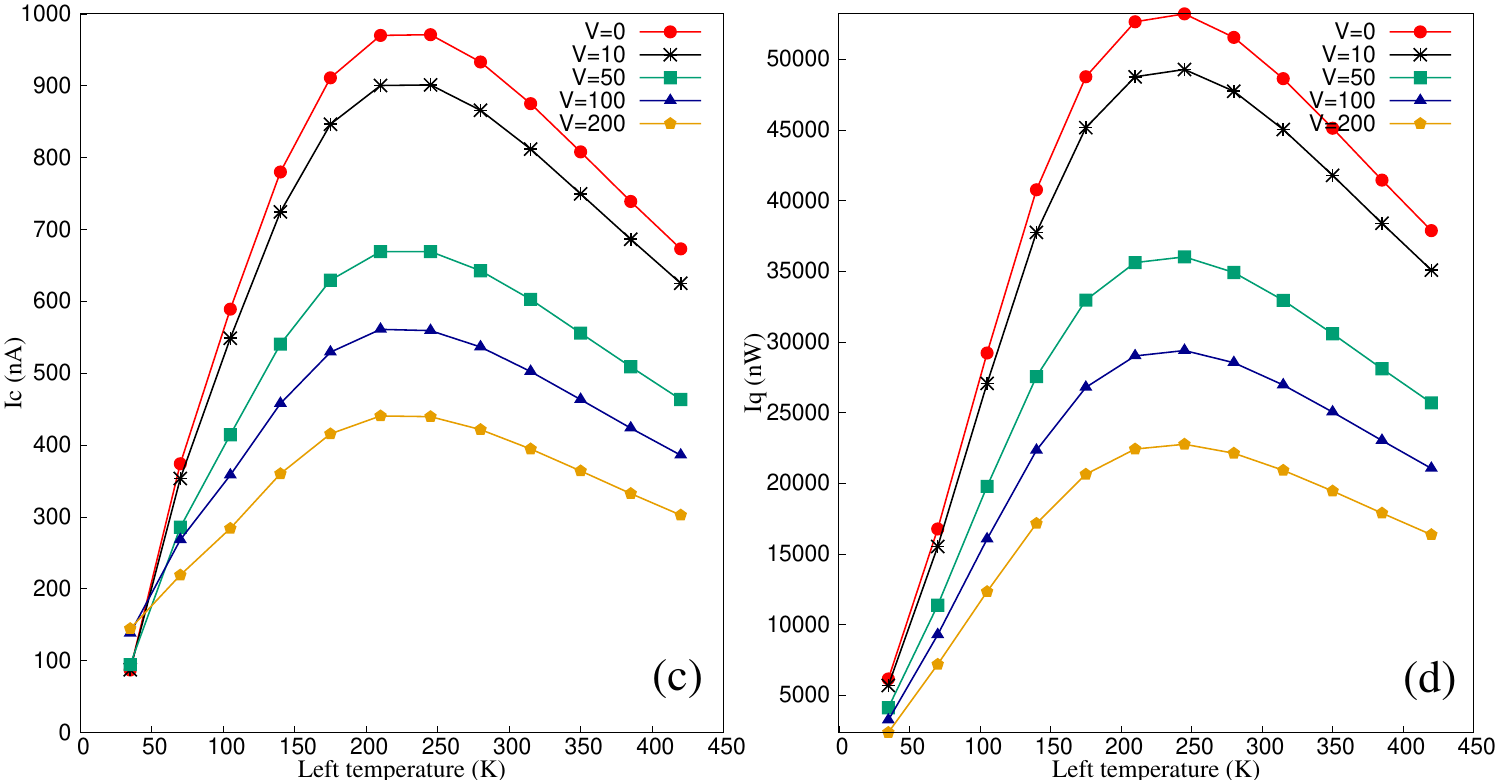}
 \caption{ Effect of different strengths of the potential associated to the impurities, shown in the legend, on the conduction of triangle shells. (a) shows the electrical current as a function of left chemical potential, and (b) shows the heat current as a function of left chemical potential. (c) shows the electrical current as function of the left temperature, and (d) shows the heat current as function of the left temperature.}
\label{strength} 
\end{figure}

In this section we consider a small number of impurities (300) with the magnitudes of their associated potential varying from zero (pure, or clean shell) to 200\,meV. Such systems, i.e. with a low number of high strength impurities, are particularly interesting because they can model different physical situations. For instance, these impurities can be considered as small, but heavy ionized doping concentrations \cite{sadi2019simulation}. And in some cases, the scattering of free carriers by phonons can be modelled  with such strong potentials \cite{markussen2007scaling,fonoberov2006giant}. 

\begin{figure}%[H] 
\centering
\includegraphics[width=1\linewidth]{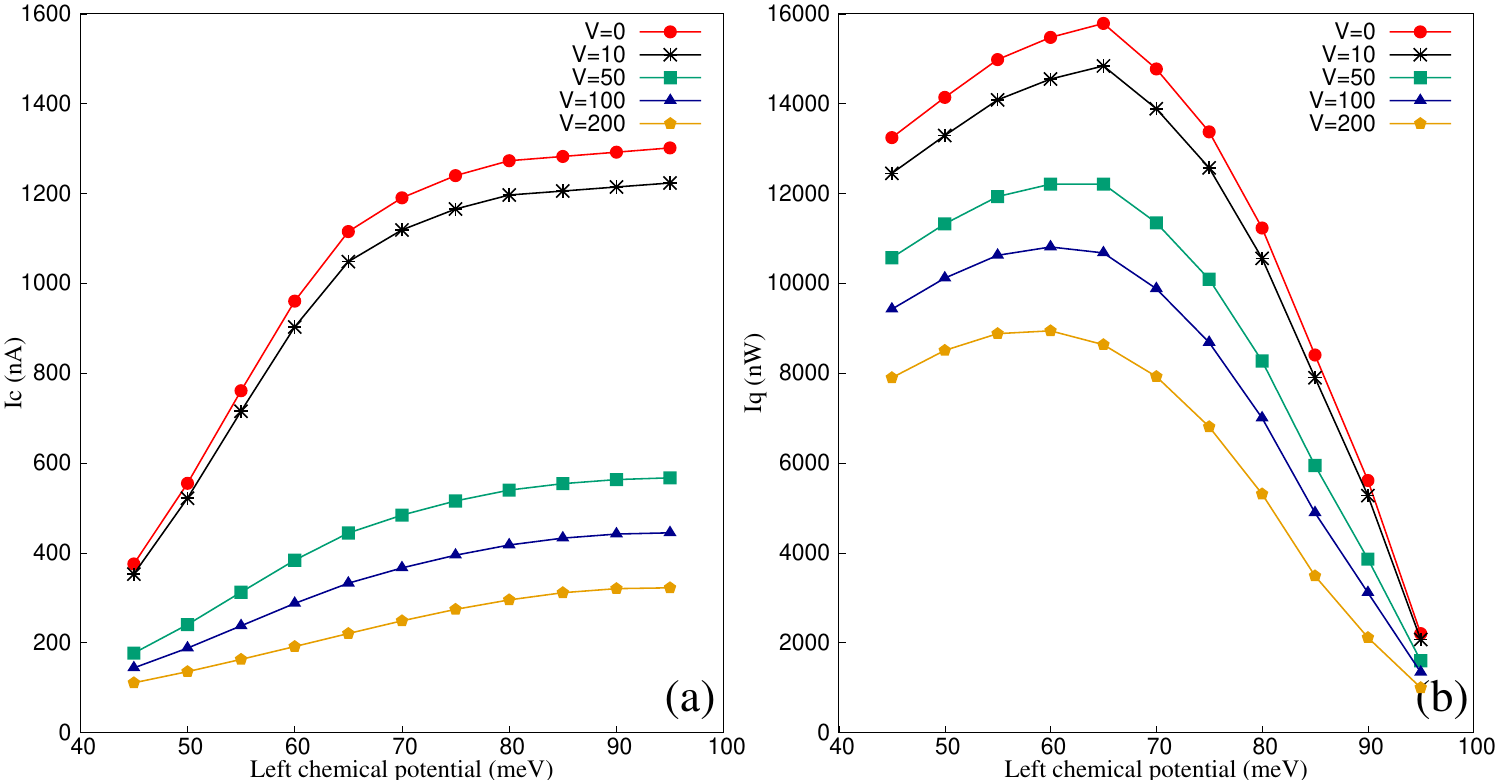}
\includegraphics[width=1\linewidth]{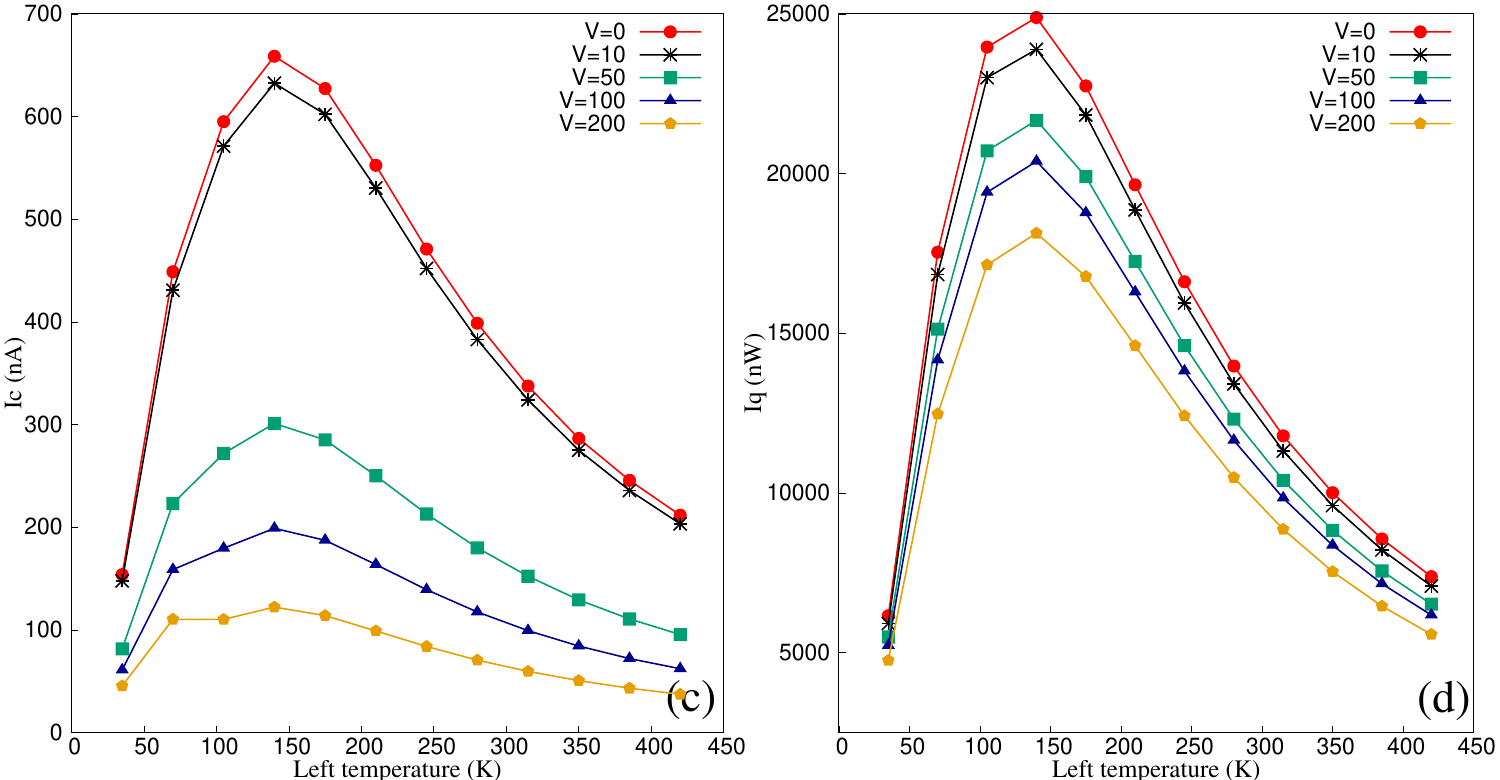}
 \caption{Effect of different strengths of the impurity potentials on the conduction of hexagonal shells. (a) and (c) figures show electrical current, and (b) and (d) show the
heat current. The conduction features are studied as a function of left chemical potential
in (a) and (b), and as a function of left temperature in (c) and (d).
}
\label{strength-hex} 
\end{figure}

Contrary to the case of large number of low strength impurities,  in the case of the small number of impurities with high strength there is a large deviation from the clean shell for all studied cases (Figure~\ref{strength} and Figure~\ref{strength-hex}). 
The reduction of the currents becomes 50\% for triangle shell and 85\% in some cases for hexagonal shells.
In Figure~\ref{strength}(a)  we show the variation of the currents with the increasing the left chemical potential.  By shifting the left chemical potential values we increase the transmission window, and thus allow, more states to participate to the transport, which leads to a higher electrical current. This increase in values of the electrical current of triangle shells with no impurities reaches almost ninefold of the initial value, while in shells with impurities this increase is strongly suppressed.  

In Figure~\ref{strength} (b) we show the heat current as a function of the left chemical potential for a few values of impurity strength. The current  decrease is significant and reaches  12-40\% in the peaks for stronger impurities. Both electrical and heat currents, as functions of left temperature, also decrease in presence of stronger impurities, Figure~\ref{strength} (c) and (d).  A close look at the values of $I_{c}$ and $I_{q}$ in Figure~\ref{strength} (c) and (d), with respect to the shell with no impurity indicates interesting features. Both the charge and the heat currents show peaks between 200-250\,K. And in this temperature range, with strong impurities in the system, we  obtain 50\% reduction of $I_{c}$, while for the same temperature range the reduction of $I_{q}$ reaches 65-70\%.
So in the presence of a temperature gradient, in the triangular shells, with strong impurities, we can reach a system that suppresses electrical current less than heat current. This feature is very desirable in thermoelectric applications.

In hexagonal shells, the charge current in the presence of intense impurities shows a significant reduction, of $\approx 75-85\%$ 
on the average, when the chemical potential and 
temperature are varied. In all cases the charge current in the hexagonal shell is much stronger suppressed than in the triangular shell. While heat current in hexagonal shells shows a smaller reduction in comparison to triangle shells. By comparing Figure~\ref{strength} and Figure~\ref{strength-hex} one can see that strong impurities lead to different behavior of the charge and heat current of hexagonal and triangular shells. For the triangular shell, the heat current is more suppressed in presence of intense impurities than the charge current, while the opposite occurs in the case of a hexagonal shell. By considering values of charge and heat current in hexagonal shells (Figure~\ref{strength-hex}) in the presence of intense impurities, this shell is a good candidate for removing and expunging heat.

\section{Conclusions}

We studied the effect of the number and strength of impurities on the charge and heat currents driven by a temperature or chemical potential bias, for different cross-sectional geometries of tubular nanowires. 
We began with a clean shell (or pure, i.\ e. without impurities) in each case and increased gradually the number of impurities.  
As expected these impurities lead to an increase of the scattering of electrons, and due to this reason the charge and heat current transported by carriers always decrease. 
%as expected these impurities lead to an increase in the scattering of electrons, and these scattered electrons do not carry charge or heat directly from one side to another which lead to a reduction in overall conduction of electrons. 
We could obtain a further reduction of the currents by using a smaller number of impurities, but with a stronger associated potential.  This 
effect may be also expected, since in general the impurity effects scale linearly with the impurity density, but quadratically with their strength. 
%{\red [SIE: This is expected since the impurity effect scales like $n_{imp} V^2$, right?]}. 
%{\blue Our result confirms $n_{imp} V^2$- but I dont know where this equation is coming from, one can solve "impurity Scattering in Semiconductors" through collision terms in the Lorentz-Boltzmann equations-but I really do not know how to derive $n_{imp} V^2$ from there. {\red Andrei can you help with this?} } 
However, we show that the charge and heat current, and the effect of impurities on them, depend on the shell geometry. In the presence of impurities the triangular shell carries a charge current almost four times more than the hexagonal shell, for the same temperature gradient (35\,K). While the heat current for the triangular shell is only 15\% higher than hexagonal shells.  An interesting result is that the effect of impurities on the charge and heat transport is smaller in the triangular shell than in the hexagonal shell.  The reason is that the localization of the electrons on the corner and on the sides of the polygonal shell is more pronounced in the triangular case, which leads to higher energy intervals between different transverse states and consequently to a reduced scattering, compared to the hexagonal case. 

{\blue In conclusion, the interplay of the geometry and impurities leads to complex effects that may favor an increase of the thermoelectric efficiency of semiconductor based tubular or core-shell nanowires, that needs to be experimentally investigated.}
 %The shells with triangle geometry have the highest electronic charge transport ability, and even with reduction of electrical current due to the presence of impurities still remain high in comparison to other geometries.

\section*{Acknowledgment}
This work was supported by the Icelandic Research Fund, Grant 195943-051 and 229078-051.

\bibliographystyle{apsrev4-1}
\bibliography{Bibliography2}

\end{document}